\documentclass[aps, prl, twocolumn, superscriptaddress, showpacs]{revtex4}

\usepackage{amsfonts,amssymb,amsmath,amsbsy}
\usepackage{epsfig}

\begin{document}
\title{Clustering of solutions in the random satisfiability problem.}


\author{M. M\'ezard}
\affiliation{Laboratoire de Physique Th\'eorique et Mod\`eles Statistiques,
b\^atiment 100, Universit\'e Paris-Sud, F--91405 Orsay, France.}
\author{T. Mora}
\affiliation{Laboratoire de Physique Th\'eorique et Mod\`eles Statistiques,
b\^atiment 100, Universit\'e Paris-Sud, F--91405 Orsay, France.}
\author{R. Zecchina}
\affiliation{Abdus Salam International Center for Theoretical Physics,
Strada Costiera 11, 34100 Trieste, Italy}

\date{\today}

\begin{abstract}

Using elementary rigorous methods we prove the existence of a clustered phase
in the random $K$-SAT problem, for $K\geq 8$. In this phase the solutions are grouped into
clusters which are far away from each other. The results are in agreement with
previous predictions of the cavity method and give a rigorous confirmation to
one of its main building blocks. It can be generalized to other systems of
both physical and computational interest.

\end{abstract}

\maketitle

Constraint satisfaction problems (CSPs) provide one of the main building blocks
for complex systems studied in computer science, information theory and
statistical physics, and may even turn out to be important in the 
statistical studies of biological networks. Typically, they involve a large
number of discrete variables, each one taking a finite number of values, and a
set of constraints: each constraint involves a few variables, and forbids some
of their joint assignments. A simple example is the $q$-coloring of a graph,
where one should assign to each vertex of the graph a color in
$\{1,\dots,q\}$, in such a way that two vertices related by an edge have
different colors. In the case $q=2$, this is nothing but the zero temperature
limit of an antiferromagnetic problem, which is known to display a spin-glass
behaviour when the graph is frustrated and disordered. CSPs also appear naturally in the studies
of structural glasses \cite{sellittoetal} and rigidity percolation \cite{barreetal}.

Given an instance of a CSP, one wants to know whether there exists a solution,
that is an assignment of the variables which satisfies all the constraints
(e.g. a proper coloring). When it exists the instance is called SAT, and one
wants to find a solution. Most of the interesting CSPs are NP-complete: in the
worst case the number of operations needed to decide whether an instance is
SAT or not is expected to grow exponentially with the number of variables. But
recent years have seen an upsurge of interest in the theory of typical-case
complexity, where one tries to identify random ensembles of CSPs which are hard
to solve, and the reason for this difficulty. Random ensembles of CSPs are
also of great theoretical and practical importance in communication theory:
some of the best error correcting codes (the so-called low density parity
check codes) are based on such constructions \cite{gallagher,mckaybook}.

The archetypical example of CSP is Satisfiability (SAT). This is a core
problem in computational complexity: it is the first one to have been shown
NP-complete \cite{cook}, and since then thousands of problems have been shown
to be computationally equivalent to it. Yet it is not so easy to find
difficult instances. The main ensemble which has been used for this goal is
the random $K$-satisfiability ($K$-SAT) ensemble. The variables are $N$ binary
variables ---\,Ising spins\,--- $\vec\sigma=\{\sigma_i\}\in\{-1,1\}^{N}$. The
constraints are called $K$-clauses. Each of them involves $K$ distinct spin
variables, randomly chosen with uniform distribution, and it forbids one
configuration of these spins, randomly chosen among the $2^K$ possible ones. A
set of $M$ clauses defines the problem. This corresponds to generating a
random logical formula in conjunctive normal form, which is a very generic
problem appearing in logic. $K$-SAT can also be written as the problem of
minimizing a spin-glass-like energy function which counts the number of
violated clauses and in this respect random $K$-SAT is seen as a prototypical
diluted spin-glass \cite{Monasson-Zecchina}. Here we shall keep to the most
interesting case $K\ge 3$ (for $K=2$ the problem is polynomial).

In the recent years random $K$-SAT has attracted much interest in computer 
science and in statistical physics 
\cite{AI,TCS,Selman-Kirkpatrick,Nature}. The interesting limit is
 the thermodynamic limit $N \to\infty,\  M \to \infty$  
at fixed clause density $\alpha=M/N$. Its most striking feature is
certainly its sharp threshold. 
 It is strongly believed that there exists a phase transition for this
problem:  Numerical and heuristic analytical
arguments are in support of the so-called
 \emph{Satisfiability Threshold Conjecture}:\\
{\it  There exists $\alpha_c(K)$ such that with high probability:\\
-- if $\alpha<\alpha_c(K)$, a random instance is satisfiable ; \\
-- if $\alpha>\alpha_c(K)$, a random instance  is unsatisfiable. \\}
In all this paper, `with high probability' (w.h.p.) means with a probability 
going to one in the $N \to \infty$ limit. Although this conjecture remains 
unproven, Friedgut has come close to it by establishing
 the existence of a non-uniform sharp threshold
\cite{Friedgut}. A lot of efforts have been devoted to understanding this
phase transition. This is interesting both from the physics point of view,
but also from the computer science one, because the random instances with $\alpha$
close to $\alpha_c$ are the hardest to solve. The most important rigorous
results so far are bounds for the threshold $\alpha_c(K)$. The best upper
bounds  were derived using first moment methods
\cite{kirousis,dubois}. Lower bounds can be found by
analyzing some algorithms which find SAT assignments \cite{franco,frieze-suen}, but recently a
new method, based on second moment methods, has found better and
algorithm-independent lower bounds \cite{achliomoore,achlioperes}. Using these bounds, it
was shown that $\alpha_c(K)$ scales as $2^K \ln(2)$ when $K\to\infty$.

On the other hand, the cavity method, which is a powerful tool from the
statistical physics of disordered systems \cite{Cavity}, is claimed to be able
to compute the exact value of the threshold \cite{MPZ,MZ,MMZ-RSA}, giving for
instance $\alpha_c(3)\simeq 4.2667...$ It is a non-rigorous method but the
self-consistency of its results have been checked by a `stability analysis'
\cite{montanariricci,monparric,MMZ-RSA}, and it also leads to the development
of a new algorithmic strategy, `survey propagation', which can solve very
large instances at clause densities which are very close to the threshold
(e.g. $N=10^6$ and $\alpha=4.25$).

The main hypothesis on which the cavity analysis of random $K$-satisfiability
relies is the existence, in a region of clause density $[\alpha_d,\alpha_c]$
close to the threshold, of an intermediate phase called the `hard-SAT' phase. 
In this phase the set $\cal S$ of 
solutions (a subset of the vertices in the$N$-dimensional hypercube)
 is supposed to split into many disconnected
\emph{clusters} ${\cal S} = {\cal S}_1 \cup
{\cal S}_2\cup\dots$. If one considers two solutions $X,Y$ in the same
cluster ${\cal S}_j$, it is possible to walk from $X$ to $Y$ (staying in $\cal
S$) by flipping at each step a finite numbers of spins. If on the other hand
$X$ and $Y$ are in different clusters, in order to walk from $X$ to $Y$
(staying in $\cal S$), at least one step will involve an extensive number
(\emph{i.e.} $\propto N$) of spin flips.
This clustered  phase is held responsible for entrapping many local search
algorithms into non-optimal metastable states \cite{montanarisemerjian}. This
phenomenon is not exclusive to random $K$-SAT. It is also predicted to appear in 
many other hard satisfiability and optimization
problems such as Coloring \cite{mulet,braunstein} or the Multi-Index Matching
Problem \cite{martinmezardrivoire}, and corresponds to a `one step replica
symmetry breaking' (1RSB) phase in the language of statistical physics. It is
also a crucial limiting feature for decoding algorithms in some  error correcting 
codes~\cite{montanari}. 
So far, the only CSP
for which the existence of the clustering phase has been established rigorously  is the simple
polynomial problem
of random XOR-SAT \cite{XORSAT-MRZ,XORSAT-CDMM}. In other cases it is an hypothesis, 
the self-consistency of which is checked by the cavity method.

In this paper we provide rigorous arguments which show the existence
of the clustering phenomenon in random $K$-SAT, for large
enough $K$, in some region of $\alpha$ included in the interval
$[\alpha_d(K),\alpha_c(K)]$ predicted by the statistical physics
analysis. Our result is not able to confirm all the details of this
analysis but it provides strong evidence in favour of its validity.

Given an instance $F$ of random $K$-satisfiability, we define a SAT-$x$-pair as  a pair of
assignments $(\vec \sigma,\vec \tau)\in\{-1,1\}^{2N}$, which both
satisfy $F$, and which are at a Hamming distance
$ d_{\sigma\tau}\equiv\sum_{i=1}^{N}(1-\sigma_i\tau_i)/2$ specified by $x$ as follows:
\begin{equation}
d_{\sigma\tau}\in [Nx-\epsilon(N),Nx+\epsilon(N)]
\end{equation}
Here $x$ is the normalized distance between the two
configurations, which we keep fixed as $N$ and $d$ go to infinity.
The  resolution $\epsilon(N)$ must be such that $\lim_{N \to \infty} \epsilon(N)/N=0$,
but its precise form is
 unimportant for our large $N$ analysis. One can choose for instance $\epsilon(N)=\sqrt{N}$.
 
We call \emph{$x$-satisfiable} a formula for which such a pair of
solutions exists.
Our study mimicks the usual steps which are taken in rigorous studies of
$K$-SAT, but taking pairs of assignments at a fixed distance
instead of single assignments.

\begin{figure} 
\begin{center}
\epsfig{file=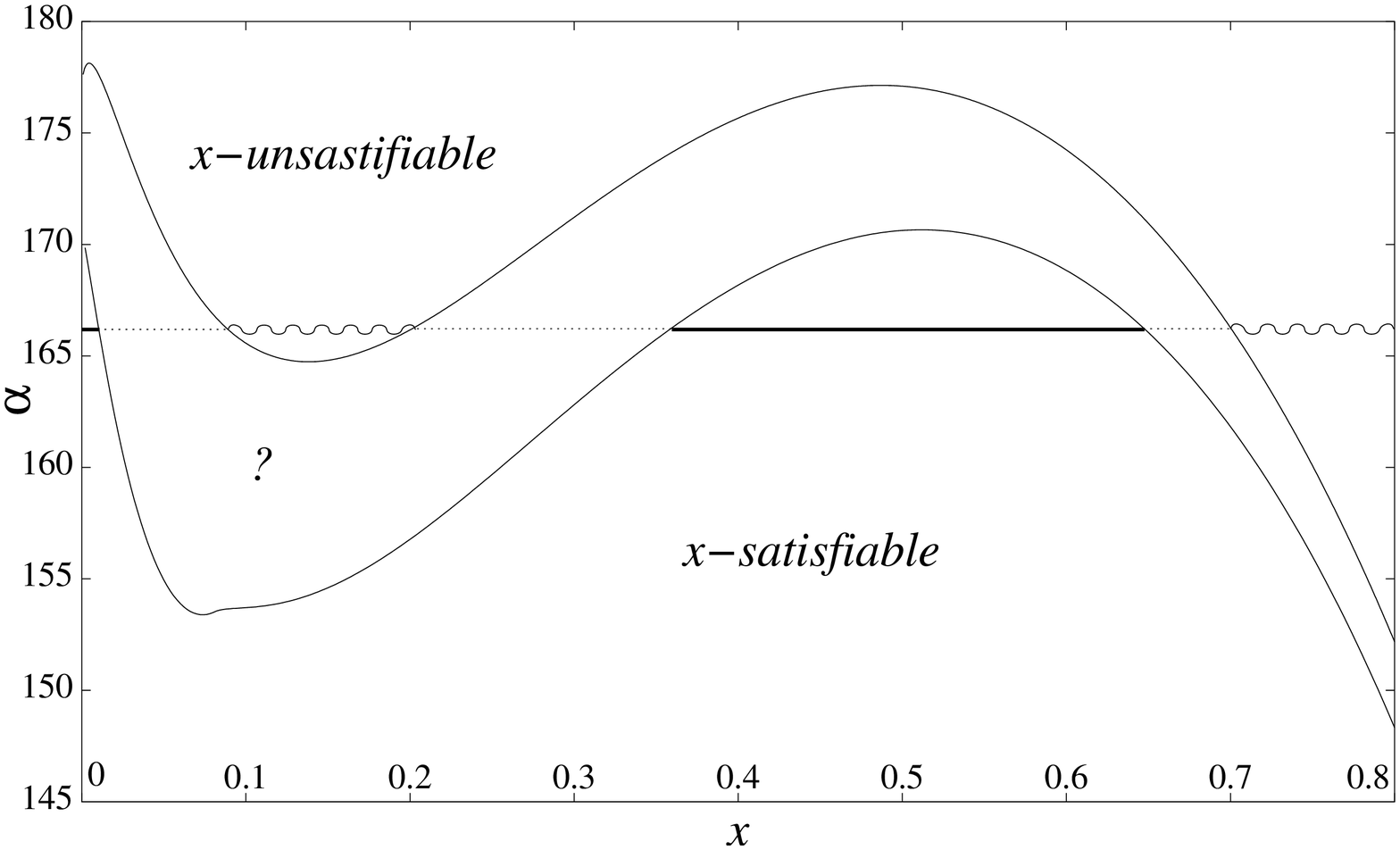,width=8cm,angle=-0} \caption{\label{alpha8} \small
Lower and Upper Bounds for the $x$-satisfiability threshold $\alpha_c(K=8,x)$.
The upper curve is obtained by the first moment method. Above this curve there
exists no SAT-$x$-pair, w.h.p.. The lower curve is obtained by the second
moment method. Below this curve there exists a SAT-$x$-pair w.h.p.. For values
of $\alpha$ lying between $164.735$ and $170.657$, these bounds guarantee the
existence of a clustering phenomenon. 
The horizontal line gives an example of this phenomenon for $\alpha=166.1$. 
 We exhibit
the successive phases as one varies $x$: $x$-SAT regions
are represented by a thick solid line, $x$-UNSAT regions by a wavy line, and
``don't know'' regions by a dotted line. The $x$-SAT region near $x=0$
corresponds to intra-cluster pairs, whereas the $x$-SAT region around $x=.5$
corresponds to inter-cluster pairs. In this example, the intermediate $x$-UNSAT region 
around $x \sim .13$ shows
 the existence of a ``gap'' between clusters. We recall that the
best refined lower and upper bounds for the satisfiability threshold
$\alpha_c(K=8)$ from \cite{dubois,achlioperes} are respectively $173.253$ and
$176.596$. The cavity prediction is $176.543$ \cite{MMZ-RSA}. }
\end{center}

\end{figure}

We first formulate the \emph{$x$-Satisfiability Threshold Conjecture}:\\
{\it For all $K\geq 2$ and for all $x$, $0<x<1$, there exists an
$\alpha_c(K,x)$ such that w.h.p.:\\ -- if $\alpha<\alpha_c(K,x)$, a
random $K$-CNF is $x$-satisfiable; \\ -- if $\alpha>\alpha_c(K,x)$, a
random $K$-CNF is $x$-unsatisfiable,\\} which generalizes the usual
satisfiability threshold conjecture (obtained for $x=0$).  We shall find
explicitly below two functions, $\alpha_{LB}(K,x)$ and $\alpha_{UB}(K,x)$
which give lower and upper bounds for $\alpha$ for $x$-satisfiability at
a given value of $K$.
Numerical computations of
these bounds show that $\alpha(K,x)$ is \emph{non monotonous} as a function of
$x$ for $K\geq 8$, as illustrated in Fig.\ref{alpha8}.  This in turn
shows that, for $K$ large enough and in some well chosen interval of
$\alpha$ below the satisfiability threshold, SAT-$x$-pairs exist for $x$
close to $0$ ($\vec\sigma$ and $\vec\tau$ in the same cluster) and $x$ close to $.5$ ($\vec\sigma$ and $\vec\tau$ in different clusters), but there is an intermediate $x$
region where they do not exist. Fig.\ref{alpha8} shows an explicit example of this scenario for a particular value of $\alpha$.

In what follows we first establish a rigorous and explicit upper bound
using a simple first moment method.  Subsequently, we provide a
(numerical) lower bound using a second moment method
\cite{achliomoore,achlioperes}.  Both results are based on elementary
probabilistic techniques which could be generalized to other physical
systems or random combinatorial problems.

{\it Upper bound: the first moment method.}  We use the fact that,
when $Z$ is a non-negative random variable:
\begin{equation}
\mathbf{P}(Z \geq 1)\leq \mathbf{E}(Z)\ .
\label{lemfirst}
\end{equation}
Given a formula $F$, we take $Z(F)$ to be the number of pairs of solutions at fixed distance (with
resolution $\epsilon(N)$):
\begin{equation}\label{Z1}
Z(F)=\sum_{\vec \sigma,\vec \tau}\delta\left(\frac{d_{\sigma\tau}}{N}\simeq x \right)
\;\delta(\vec\sigma, \vec\tau \in S(F)) \; \ ,
\end{equation}
where
$S(F)$ is the set of solutions to $F$.  Throughout this 
paper $\delta(A)$ is an indicator function, equal to $1$ if the
statement $A$ is true, and to $0$ otherwise.  Since $Z(F)\geq 1$
is equivalent to ``$F$ is $x$-satisfiable'', (\ref{Z1}) gives an upper
bound for the probability of $x$-satisfiability.
The expected value of the double sum over the  choice of a random $F$ is:
\begin{equation}
\mathbf{E}(Z(F))=2^N\binom{N}{Nx}{\mathbf{E}\left[\delta(\vec\sigma,\vec\tau\in S(c))\right]}^{M}.
\end{equation}
We have used  $\delta(\vec\sigma,\vec\tau\in S(F))=\prod_c
\delta(\vec\sigma,\vec\tau\in S(c))$, where $c$ denotes  the
 clauses, and the fact that  clauses are drawn independently.
The expectation
$\mathbf{E}\left[\delta(\vec\sigma,\vec\tau\in S(c))\right]$ is equal to: $1-2^{1-K}+2^{-K}(1-x)^{K}$ (there
are only two realizations of the clause among $2^K$ that do not
satisfy $c$ unless the two configurations overlap exactly on the
domain of $c$).

In the thermodynamic limit, $\ln \mathbf{E}(Z(F))/N \to \Phi_1(x,\alpha) $,
where: 
\begin{eqnarray}
\Phi_1(x,\alpha)
=\ln 2+H_2(x) +\alpha\ln\left[1-2^{-K}(2-(1-x)^{K})\right],
\nonumber
\end{eqnarray}
where $H_2(x)=-x\ln x-(1-x)\ln(1-x)$ is the two-state entropy
function. This gives the upper bound:
\begin{equation}
\alpha_{UB}(K,x)=-\frac{\ln 2+H_2(x)}{\ln(1-2^{1-K}+2^{-K}(1-x)^{K})}.
\end{equation}

{\it Lower bound: the second moment method. } We use the
fact that, when $Z$ is a non-negative random variable:
\begin{equation}
\mathbf{P}(Z>0)\geq \frac{\mathbf{E}(Z)^2}{\mathbf{E}(Z^2)}.
\end{equation}
However using this formula with $Z$ equal to the number of solutions
fails, and one must instead use a weighted sum \cite{achliomoore}.
We follow the strategy recently developed in \cite{achlioperes}, which we generalize 
 to SAT-$x$-pairs by taking:
\begin{equation}\label{Zw}
Z(F)=\sum_{\vec\sigma,\vec\tau}\delta\left(\frac{d_{\sigma\tau}}{N}\simeq x\right)
 \prod_c W(\vec\sigma,\vec\tau,c)\ .
\end{equation}
$W(\vec\sigma,\vec\tau,c)$ is a weight associated with the clause $c$, given the couple $(\vec\sigma,\vec \tau)$, and is defined as follows: Suppose that $c$ is satisfied by $n_\sigma$ among the $K$ $\vec\sigma$-variables involved in $c$, and by $n_\tau$ among the $K$ $\vec\tau$-variables. Call $n_0$ the number of common values between the $\vec\sigma$- and $\vec\tau$-variables involved in $c$. Then define:
\begin{equation}
W(\vec\sigma,\vec\tau,c)=\left \{\begin{array}{ll}\lambda^{n_\sigma+n_\tau}\nu^{n_0}&\textrm{if }n_\sigma>0\textrm{ and }n_\tau>0,\\
0& \textrm{otherwise.}\end{array}\right.
\label{Wdef}
\end{equation}
Note that with this definition of $Z$, the choice $\lambda=1,\nu=1$ simply yields the number of solutions (\ref{Z1}).

Let us now compute the first two moments of $Z$ (\cite{MMZ_inprep}):
\begin{equation}
\mathbf{E}(Z)=2^N\binom{N}{Nx}\left[f_1^{(\lambda,\nu)}(x)\right]^M \ ,
\end{equation}
where $f_1^{(\lambda,\nu)}(x)=\mathbf{E} (W(\vec\sigma,\vec\tau,c))$ can be calculated by simple combinatorics (via multinomial sums). To compute $\mathbf{E}(Z^2)$, we sum over  four spin configurations $\vec\sigma,\vec\tau,\vec\sigma',\vec\tau'$. Symmetry allows to fix $\sigma_i=1$. Let $Na(t,s,t')$ be the number of sites $i$ such that $\tau_i=t$, $\sigma_i'=s'$ and $\tau_i'=t'$ (where $t,s,t' \in\{\pm 1\}$). It turns out that the term of the sum depends only on these 8 numbers $a(\pm 1,\pm 1,\pm 1)$. We collect them into a vector $\mathbf{a}$ and get:
\begin{equation}\label{E2}
\mathbf{E}(Z^2)=2^N \int_V d \mathbf{a} \; \frac{N!}{\prod_{t,s',t'}(Na(t,s',t'))!}
\left[f_2^{(\lambda,\nu)}(\mathbf{a})\right]^M \ ,
\end{equation}
where $f_2^{(\lambda,\nu)}(\mathbf{a})=\mathbf{E}(W(\vec\sigma,\vec\tau,c)W(\vec\sigma',\vec\tau',c))$ can
be calculated by simple combinatorics
in the same way as $f_1$. The integration set $V$ is a 5-dimensional simplex taking into account the normalization $\sum_{t,s',t'} a(t,s',t')=1$ and the two constraints: $d_{\sigma\tau}/N\simeq x$, $d_{\sigma'\tau'}/N\simeq x$.

A saddle point evaluation of eq.(\ref{E2}) gives, for $N \to \infty$:
\begin{equation}\label{cond}
\frac{\mathbf{E}(Z)^2}{\mathbf{E}(Z^2)}\geq C_0\exp(-N\max_{\mathbf{a}\in V}\Phi_2(\mathbf{a})),
\end{equation}
where $C_0$ is a constant depending on $K$ and $x$, and:
\begin{equation}\label{defphi}
\Phi(\mathbf{a})=H_8(\mathbf{a})-\ln 2-2H_2(x)+\alpha  \ln f_2^{(\lambda,\nu)}(\mathbf{a})-2 \alpha \ln f_1^{(\lambda,\nu)}(x),
\end{equation}
with $H_8(\mathbf{a})=-\sum_{t,s',t'} a(t,s',t') \ln a(t,s',t')$. In general $\max_{\mathbf{a}\in V}\Phi(\mathbf{a})$ is non-negative
and one must choose appropriate weights
 $W(\vec \sigma,\vec \tau,c)$ in such a way that
$\max_{\mathbf{a}\in V}\Phi(\mathbf{a})= 0$.
We notice that at the particular point $\mathbf{a}^*$ where $(\vec\sigma,\vec\tau)$ is uncorrelated with $(\vec\sigma',\vec\tau')$, we have $\Phi(\mathbf{a}^*)=0$.
We  fix the  parameters $\lambda$ and $\mu$ defining the weights (\ref{Wdef})  
in such a way that $\mathbf{a}^*$
be a local maximum of $\Phi$. This gives two algebraic equations in $\lambda$ and $\nu$
which  have a unique
solution $\lambda>0,$ $\nu>0$. Fixing
$\lambda$ and $\nu$ to these values, $\alpha_{LB}$ is the largest value of
$\alpha$ such that the local maximum at $\mathbf{a}^*$ is a \emph{global} maximum,
i.e. such that there exists no $\mathbf{a}\in V$ with $\Phi(\mathbf{a})>0$:
\begin{equation}\label{alpha2}
\alpha_{LB}(K,x)=\inf_{\mathbf{a}\in V}\frac{\ln
  2+2H_2(x)-H_8(\mathbf{a})}{\ln f_2^{(\lambda,\nu)}(\mathbf{a})-2\ln f_1^{(\lambda,\nu)}(x)},
\end{equation}

We devised several numerical 
strategies to evaluate
$\alpha_{LB}(K,x)$. The implementation of Powell's method starting from each
point of a grid of size $\mathcal{N}^5$ ($\mathcal{N}=10,15,20$) on
$V$ turned out to be the most efficient and reliable.  The results are
given by Fig.\ref{alpha8} for $K=8$, the smallest $K$ such that the
clustering conjecture is confirmed. We found a clustering phenomenon
for all the values of $K \ge 8$ that we checked, and in fact the 
relative difference $[\alpha_{UB}(K,x)-\alpha_{LB}(K,x)]/ \alpha_{LB}(K,x)$
seems to go to zero at large $K$. 

We have shown a simple probabilistic argument which shows rigorously the existence of 
a clustered `hard-SAT' phase. The prediction from the cavity method 
is in fact a weaker statement. It can be stated 
 in terms of the overlap distribution function $P(x)$, which is the
probability, when two SAT-assignments are taken randomly (with uniform
distribution), that their distance is given by $x$. The cavity method finds that
this distribution has a support concentrated on two values: a large value
$x_1$, close to one, gives the characteristic `radius' of a cluster, a smaller
value $x_0$ gives the characteristic distance between clusters. This
 does not imply that there exists no pair of solution for values of
$x$ distinct from $x_0,x_1$: it just means that such pairs are exponentially
less numerous than the typical ones. Our rigorous result shows that in fact
there exists a true gap in $x$, with no SAT-$x$-pairs, at least for $K \ge 8$.
More sophisticated moment computations might allow to get some results for
smaller values of $K$. Still the conceptual simplicity of our computation
makes it a useful tool for proving similar phenomena in other systems of
physical or computational interests, like for instance the graph-coloring
(antiferromagnetic Potts) problem.

This work has been supported in part by the EC through the network 
MTR 2002-00319 `STIPCO' and the FP6 IST consortium `EVERGROW'.

\end{document}